# Photonic unsupervised learning processor for secure and high-throughput optical fiber communication

Yitong Chen[1,3]*, Tiankuang Zhou[1,3]*, Jiamin Wu[1,3,4], Hui Qiao[1,3], Xing Lin[2,3,4,†], Lu Fang[2,3,4,†] and Qionghai Dai[1,3,4,†]

[1]Department of Automation, Tsinghua University; Beijing 100084, China
[2]Department of Electronic Engineering, Tsinghua University; Beijing 100084, China
[3]Institute for Brain and Cognitive Sciences, Tsinghua University; Beijing 100084, China
[4]Beijing National Research Center for Information Science and Technology, Tsinghua University; Beijing 100084, China
*These authors contributed equally to this work
†Corresponding author. Email: qhdai@tsinghua.edu.cn

**Following the explosive growth of global data, there is an ever-increasing demand for high-throughput optical fiber communication (OFC) systems to perform massive data transmission and processing. Existing OFC methods mainly rely on electronic circuits for data processing, which severely limits the communication throughput. Though considered promising for the next-generation high-speed fiber communication, all-optical OFC remains unachievable due to serious challenges in effective optical computing, system modeling and configuring. Here we propose an end-to-end photonic encoder-decoder (PED) processor which maps the physical system of OFC into an optical generative neural network. By modeling the OFC transmission process as the variation in the constructed optical latent space, the PED learns noise-resistant coding schemes via unsupervised optimization. With multi-layer parametric diffractive neural networks, the PED establishes a large-scale and high-throughput optical computing framework that integrates the main OFC computations including coding, encryption and compression to the optical domain. The whole system improves the latency of computation in OFC systems by five orders of magnitude compared with the state-of-the-art device. On benchmarking datasets, the PED experimentally achieves up to 32% reduction in transmission error ratio (ER) than on-off keying (OOK), one of the mainstream methods with the lowest ER in general transmission. As we demonstrate on medical data, the PED increases the transmission throughput by two orders of magnitude than 8-level pulse amplitude modulation (PAM-8). We believe the proposed photonic encoder-decoder processor not only paves the way to the next-generation all-optical OFC systems, but also promotes a wide range of AI-based physical system designs.**

Optical communication has been widely used in modern society, ranging from wireless telecommunication and satellites [1–4] to optical fiber-based data centers and undersea communication [5–9]. It adopts photons as the information carrier, which makes it possible to transmit large-scale data at high efficiency. Despite the dominance of electronic processing, over 95% of digital information globally is transmitted optically, constructing an optical transmission and electronic processing information society [10,11]. With the explosive growth in global data and its major requirement of transmission via optical fiber networks, classical data coding strategies base themselves mainly on electronic circuits[12], ignoring the great potential of optical computing and, hence, confronting the difficulty of increasingly generating enough data-carrying capacity [13–16]. In addition, processing the transmitted massive optical signals, such as encryption as well as compression, has placed critical burdens on current electronic computing platforms. The gap of the operating frequency between the state-of-the-art electronic circuits and the fiber transmission lines is orders of magnitude due to the limited drift velocity of electrons[13,16–19]. Therefore, all-optical OFC systems have attracted momentous interest as the next-generation of OFC to meet the urgently-growing data demand and support the continuous scaling of transmission throughput[20–22].

Engineering an OFC system fundamentally includes pre-processing (encoding) messages for encryption, anti-noise coding and compression, coupling into fiber channels and post-processing (decoding) reconstructing messages from the received signals [5,23]. Effective efforts have been made in exploring all-optical OFC systems[24–26], such as the optical code-division multiple-access system (OCDMA)[27] and optical amplifers[28]. However, the current progress is restricted in the implementation of sporadic parts, which makes conducting the whole OFC system optically remains an enormous challenge, due to the failure to integrate the main computation processes, e.g. the encoding and decoding of data, in optics.

The recent surge of intelligent photonic computing is considered promising to provide a solution to overcome the electronic bottleneck by manipulating optical signals at ultra-high speed. Diffractive neural networks[29–32], coherent nanophotonic circuits[33], convolutional accelerators[34,35], and other optoelectronic devices[36–38] successfully realize parallel photonic fully-connected and convolution neural networks and increase the operational efficiency significantly. However, the existing all-optical neural networks are mainly restricted to limited decision tasks, for example, digits classification and edge detection. Conducting more complicated tasks such as coding and compression demands more advanced optical computing architectures including unsupervised learning and generative models, as well as end-to-end optical modeling that matches the real physical system, which considerably restricts the application of optical computing in OFC.

Here we propose an unsupervised learning photonic model to map the physical system of OFC into an optically implemented generative neural network, which we term photonic encoder-decoder (PED). The end-to-end learned photonic encoder-decoder structure enables it to encode the optical input information into a compressed and encrypted optical latent space (OLS) for fiber transmission and decode the transmitted distorted signals from the encrypted domain to the reconstructed information optically. The elaborately designed diffractive neurons directly manipulate the information in the optical domain without loss and distortion caused by the convertors to increase the transmission precision. Modeling the transmission noise as the variations in the OLS and constraining it with the probability distribution entitle the PED to learn an optimal coding scheme to maximize its noise resistance. Through unsupervised learning, the

PED compresses the data with inadequate information instead of distinct labels, which is usually inaccessible in OFC.

In our prototype optical system, the PED improves the latency of computation in OFC systems by five orders of magnitude compared with the state-of-the-art CPU. On benchmarking datasets, PED experimentally achieves up to 32% reduction in transmission error ratio (ER) than on-off keying (OOK), one of the mainstream methods with the lowest ER. Additionally, the PED increases the transmission throughput by two orders of magnitude than PAM-8 and 87 times than conventional compression method discrete cosine transformation (DCT) as we demonstrate on medical data.

**Modeling of the PED**

In the proposed PED communication architecture (Fig. 1a), the input messages are encoded into an optical latent space (OLS) with an optical artificial neural network (ANN) encoder and are later correspondingly coupled into the single-mode fiber bundle. The noise during transmission is modeled as the variation in the OLS in a variational autoencoder (VAE)-based architecture[39]. An optical ANN decoder decodes the transmitted latent space representations, i.e., the combination of the Gaussian speckles after the collimating lenses, for the faithful reconstruction of input messages (see the Supplementary Information for the mathematical modeling of the mapping from the physical OFC system to the PED). The structure and parameters of the OFC system are determined during the end-to-end unsupervised training of the PED. Since the OLS highly encrypts the input messages, it prevents fiber tapping and guarantees the security of message transmission. The PED allows two modes of transmission: general and data-specific modes, as shown in Fig. 1b and Fig. 1c, respectively. The general mode provides fundamental secure and noise-resistant transmission for arbitrary information optically. When some prior information is obtained, not necessarily labels, the data-specific mode additionally allows significant improvement in throughput.

Figure 2a is the system design of the all-optical PED, establishing the encoding-transmission-decoding flow without the requirement of any analog-digital/digital-analog (AD/DA) and opt-electrical conversions by carrying out all processes in the analog optical field, which significantly improves the speed and efficiency of communication. As depicted in Fig. 2a, after an all-optical encoder composed of reconfigurable diffractive layers, the PED passively couples the light field into a fiber bundle through a lens array and decodes the OLS with an all-optical decoder back to the original data. Every lens in the lens array couples one corresponding area of the optical field to one single-mode fiber in the fiber link. The coupling efficiency depends on the light field and the characteristics of the fibers and lens. The light field in the fiber can be considered as a linear superposition of different spatial frequencies with respective coupling coefficients (see Methods for the modeling of the coupling process). As a result, the PED encodes the information into both the amplitude and phase domains during the transmission in fibers, which achieves better results in both reconstruction and noise resistance.

The latent space of the VAE architecture is a hyperspace representation of inputs that is subject to a continuous distribution function[39]. In the PED, we generate and evaluate the OLS by dividing the output plane of the encoder into subregions and sum the low-frequency component of the optical field at each subregion by coupling it into a single-mode optical fiber. Different subregions correspond to different single-mode fibers of the fiber bundle, which can avoid the cross-talks of the spatial modes compared with transmission via a multimode fiber. The number of subregions, which corresponds to the number of optical fibers for spatial multiplexing, determines the dimensionality of the OLS. After fiber transmission, the OLS is collimated into the optical decoder to reconstruct the inputs.

In the general mode, Fig. 2b shows an example of the PED transmission results under a signal-to-noise ratio (SNR) of 9 dB, i.e., corresponding to a transmission distance of ~1600 km (see the Supplementary Information for the modeling of the distance and SNR). The input data is coded into binary blocks as bits (the first column) and the transmission noise deteriorates the input bit codes and generates incorrect bit codes after binarization (the second and third columns). During the PED-based communication, messages are encrypted by the encoder into the OLS as a complex optical field for transmission, where the same amount of transmission noise is introduced. The intensity of the optical field and its binarized result are shown in the fourth and fifth columns of Fig. 2b. By modeling the transmission noise as the variation of the OLS during the unsupervised training of the PED, the decoder reconstructs the correct bit code from a distorted OLS (Fig. 2b, the sixth and seventh columns).

Further, the OLS allows unsupervised compression when some prior information of the transmission messages are obtained, which is common in data centers and factories. Figure 2c demonstrates an example in the data-specific mode over fashion products, i.e., the Fashion-MNIST[40]. The PED encoder maps the 28×28 input image (a shirt here) into a hyper-point $L_0$ in a 36-dimensional OLS and the decoder reconstructs the input faithfully from the compressed and encrypted OLS (see the complete inputs, outputs and masks in Fig. S1). The variation during training allows the PED to obtain a continuous and noise-resistant OLS as depicted in Fig. 2c. We interpolate between different hyper-points in the OLS, such as $L_{-1}$, $L_0$ and $L_1$, to which the PED encoder maps a pair of trousers, a shirt and a sneaker respectively. The interpolation step is uniform between each image, whereas the PED outputs remain close to these fashion products around each hyper-point instead of reconstructing inexistent items. It shows that the unsupervised OLS can correctly reconstruct the items even disturbed by distinct transmission noise, implying the exceeding robustness of the PED transmission.

The compression ratio, which depends on the dimensionality of the OLS, influences the reconstruction quality as shown in Fig. 2d. The image fidelity (correlation) of the transmission outputs, averaged over the MNIST testing dataset (10k images, 28×28 grayscale pixels each) under different OLS dimensions is displayed. The results demonstrate that the transmission performance improves with the increasing OLS dimensionality rapidly when the compression ratio is huge and gradually improves when the dimensionality is adequately large. Example results of the transmitted digit "2" under different OLS dimensionalities are shown.

**General mode for secure and high-accuracy transmission**

Arbitrary data are encoded into binary bits and transmitted based on OOK in the general mode, while applying a pre-trained non-data-specific PED provides photonic encryption during transmission in fibers as the encoder embeds the information into the OLS. High precision can be achieved simultaneously with the noise-resistant coding scheme optimized by the PED (see Methods for the details of experimental modeling and set-up). Figure 3a displays the experimental results of encryption by the PED over a binary image (the logo of Tsinghua University). 3 × 3 bits are transmitted in each frame, as shown in Fig. 2b, and arranged back into the image. The information transmitted in fiber bundles, which may potentially be eavesdropped, is distorted into an unrecognizable image with the ER orders of magnitude higher than the reconstructed image (22.81%, 28.25%, 20.30%, 23.11% and 0, 0, 0, 0, relatively). The letters ('VER'), digits ('191') and Chinese characters are effectively encrypted beyond recognition. The OLS produces the coupling relationship between all latent values, which makes interception with only some of the fibers relatively impossible.

Besides effective encryption on characters, the PED shows exceeding encryption performance over natural grayscale images, such as the portrait in Fig. 3b, i.e., a 150×150 "Lena" image with 256 grayscale levels. Each grayscale pixel is encoded in one frame that has 9 binary bits. We do not change the number of bits in each frame according to the image because the grayscale depth of the data may not always be prior information under real communication situations. The PED experimentally achieves an ER of only 0.21% in reconstruction, while the ER in OLS is 89.44% on the whole image. Zoomed-in features are shown in Fig. 3b, where all details are unrecognizable, including the edge, eye, hair and mouth, with the ER encrypted to 88.54%, 95.26%, 89.63% and 92.44% and reconstructed to 0, 0.40%, 0, 0, respectively.

We also display the experimental performance of the PED over various media, such as audio (Fig. 3c). We use four classes of audios (instrument, speech, animal and vehicle) in the Google AudioSet [41], from which 20 ms each are selected and tested in a pre-trained non-data-specific PED. The input (the red line) and the PED output (the blue dots) coincide well, while the information in the OLS (the green dotted line) acutely changes. The overall ER of OLS is 88.69% (3129/3528), and the output audio reduces it to 0.14% (5/3528).

Another important purpose of coding in an OFC system is to resist transmission noise, which is realized with a PED by transmitting trained OLS. We use additive white Gaussian noise (AWGN) channels to test the performance of the PED based on different transmission distances (see the Supplementary Information for the modeling of transmission noise). Each grayscale pixel with a maximum grayscale depth of 512 is encoded into 9 bits in one frame. We train the PED over all $2^9$ instances with random noise, as shown in Fig. 3e (each line for one instance). The SNR is 9 dB for both with and without the PED, which can also be inferred from the error of the input (the second column) and the OLS (the fourth column). The threshold is grid-searched during calibration such as Fig. 3e and fixed during practical communications, such as Fig. 3d. We transmit a grayscale portrait (150×150 "Lena") and each grayscale pixel of the transmitted image is encoded into 9 bits in one frame. The variance in the OLS dexterously models the transmission noise during training, which significantly increases the anti-transmission-noise ability of the model under different SNRs, i.e., different transmission distances or systems. The PED decreases the ER by 22.17%, 32.05% and 26.22% (relatively) at 9 dB, 11 dB and 13.5 dB, respectively, compared with the OOK, one of the most noise-robust methods in existing optical communications. As the noise level increases, the effect of the PED first rises and subsequently decreases, because when the SNR is high, the experimental error of the PED dominates the results, while the PED may be unable to correct the OLS if it diverges from the ground truth too much when the SNR is quite low. Both the quantitative evaluation of the ER and the appearance of the images are distinctly improved compared with the OOK at a large range of SNRs (example images at 9 dB are shown in Fig. 3d).

**Data-specific mode for high-throughput transmission**

When applied in situations with some prior information of the transmission messages, such as the classes or features of the data, the PED allows extra compression in addition to secure transmission after unsupervised data-specific learning. Figure 4a shows an example of the PED trained to transmit the images of handwritten digits using the MNIST training set, where both the encoder and decoder are implemented with diffractive neural networks. The input digits with a size of 28×28 are transmitted by transforming into the OLS representations with a dimensionality of 6×6. Fig. 4a displays the experimental results of transmitting ten example digits, i.e., 0, 1, …, 9, from the test set (the first row). The OLS representations of the input images are generated by experimentally coupling the optical fields at different subregions of the encoder output plane with single-mode fibers and combining them. As the single-mode fiber transmits only the fundamental mode inside, it decreases the dimensionality of the transmission information. The transmission noise is modeled by setting the SNR to 24 dB with the phase noise obeying an additive Gaussian distribution (see the Supplementary Information for the modeling of noise). After the fiber transmission, the fiber outputs (the second row), are decoded by the decoder to reconstruct the inputs (the third row). We exhibit the comparisons with the discrete cosine transform (DCT) compression (the fourth row), with the reconstruction fidelities labeled on each corner. The results demonstrate that the PED system successfully encrypts the inputs for optical fiber transmission and reconstructs the input digits from their low-dimensional OLS representations at a high compression ratio. Compared with the transmission results using DCT compression, the proposed PED reconstructs input digits with significantly improved quality under the same compression ratio. The average fidelity of the experimental PED results over the MNIST testing set with 10k images is 0.81, whereas the transmission with DCT compression only achieves 0.71 (Fig. 4c). We also calculate the percentage of bad matching pixels (BMP), which is a common evaluation index for reconstructions (Fig. 4d), in which the PED outperforms DCT over both thresholds, indicating the advantage of PED in both details (smaller threshold) and outlines (larger threshold).

We further validate the PED on the transmission of a medical dataset: adrenal CT images[42]. The original binary 3D CT images are projected by one axis to generate a 28×28 grayscale 2D-image for each adrenal gland. Fig. 4e displays the experimental results of the PED over two examples from normal and pathological classes respectively, under different compression ratios. When the compression ratio is 31 and 87, the corresponding dimensionality of the OLS is 25 and 9. The PED successfully reconstructs the shape and features of the adrenal gland (the second and fourth columns) while the DCT fails to retain any details at such high compression ratios (the third and fifth columns). The reconstruction quality of the PED remains stably high even at a quite huge compression ratio (87 times) and different classes can be easily distinguished after compression and reconstruction, which is unfulfillable to DCT (the fifth column). It not only demonstrates the powerful capability in transmission and compression of the PED, but also indicates its potential to do intelligent computation such as diseases diagnosis and semantic comprehension during transmission. The fidelity of the reconstructed image after transmission by the PED, DCT, fast Fourier transform (FFT) are 0.86, 0.77, 0.43 when compression ratio is 31, 0.85, 0.69, 0.42 when compression ratio is 49 and 0.85, 0.61, 0.40 when compression ratio is 87, respectively, as depicted in Fig. 4f. The PED significantly outperforms mainstream compression methods DCT and FFT experimentally. Taking the adrenal data as an example, the PED improves the transmission throughput by two orders of magnitude than PAM-8 and ~87 times than DCT. The all-optical PED improves the computation latency by five orders of magnitude than the state-of-the-art CPU (see the Supplementary Information for calculation details).

Besides, noiseless input in OFCs represent only some situations because modern communication systems usually consist of multilevel transmission. The error accumulated during long-distance transmission and plenty of optical-electrical conversions results in noisy input when transmitted to downstream links. Therefore, we further demonstrate the PED dealing with noisy input in Fig. S3. It shows the experimental results of transmitting ten examples of MNIST handwritten digits, i.e., 0, 1, …, 9 (the first row), with additive Gaussian noise. The encoder encrypts and extracts the information before the coupling system subsamples the light field into a 6×6 OLS. The transmission noise in the fiber bundle, with the SNR of 24 dB and additive Gaussian phase noise, is modeled as the variation in OLS (the second row). The PED achieves comparable reconstruction performance (the third row) as in Fig. 4a because the lossy compression with the PED entitles the PED to powerful denoising capability. The results prove that the PED is qualified for a wide range of transmission distances and multi-level transmission.

**Discussion**

By proposing an unsupervised photonic encoder-decoder architecture, the PED achieves exceeding performance in not only the throughput and error ratio but also the efficiency and speed. It has great potential to break through the electronic bottleneck of the existing OFC by mapping the physical system to an end-to-end designed photonic neural network. Besides, by processing the original optical signal with richer information instead of distorted data after optoelectronic conversions, the PED proposes a promising direction to solve the bottleneck for conventional coding methods in transmission depth[43].

With the light-based processing system presented by the PED, it provides a photonic communication platform that facilitates the merging of other existing OFC techniques without extra conversions. For example, the PED achieves prominently higher transmission throughput and more advanced computation ability by combining with the state-of-the-art multiplexing methods such as the dense wavelength division multiplexing (DWDM)[44] and space division multiplexing[45,46] (see the Supplementary Information for details). Instead of designing a dedicated device for specific functions in OFC, the PED introduces a framework to unite existing techniques and realize computation simultaneously integrated in communication.

Compared with existing optical neural networks, the PED extends the application beyond decision-making tasks to significantly wider generative scopes. Figure S4 demonstrates applications of the PED as a universal unsupervised optical generative model in style transferring and human action video enhancement. Since decision tasks and labeled data are only a small portion in real life, the demonstration of an unsupervised generative photonic model with ultra-high speed and the ability of parallel computation will be capable of revolving plenty of application fields (see the Supplementary Information for more examples).

**Conclusion**

In this article, we propose an unsupervised photonic encoder-decoder for optical fiber communication. It not only provides an end-to-end learned processor for data processing in OFC including encoding, encryption, compression and decoding all-optically, but also improves the communication quality by processing in the optical domain directly. Two modes of the PED are introduced: the general mode for arbitrary data coding and encryption, and the data-specific mode for significantly high throughput. The transmission error ratio experimentally decreases by up to 32% by the PED as demonstrated on a benchmarking dataset. The PED widens the transmission throughput by two orders of magnitude than PAM-8 and 87 times than DCT over displayed medical data. The all-optical PED reduces the computation latency by five orders of magnitude compared with the state-of-the-art CPU.

By co-designing the encoder, decoder and fiber system all in the optical domain, our work makes inherent connections between the unsupervised learning architecture and the physical model of fiber communication systems, inspiring the next-generation all-optical communication systems with higher throughput, accuracy and data security. We believe the proposed generative photonic computing system and the end-to-end unsupervised photonic learning method will facilitate a wide range of AI applications including 6G, medical diagnosis, robotics, and edge computing.

## Methods

### PED modeling and training details

We use a VAE architecture to establish the PED, which is one of the most mainstream unsupervised generative models with various application scopes[39]. The encoder-decoder structure perfectly matches the OFC process and its ability to reconstruct data from a low-dimensional domain enables the PED to compress and encrypt the transmitted information. Additionally, the VAE provides better performance in noise resistance due to the variation in latent space when training. The loss function of the PED for optimization can be described as

$$L = \alpha l_{KLD} + \beta l_{MSE} + \gamma l_{OP},$$

where $l_{KLD}$ is the Kullback–Leibler divergence that guarantees the distribution in the optical latent space close enough to the ideal distribution, i.e., Gaussian distribution; $l_{MSE}$ is the mean square error between the PED output and the input; and $l_{OP}$ is a penalty term. To prevent the middle results from changing fiercely between the diffractive layers to make the images sick or sensitive to noise, we use an optical penalty term in the loss, similar to what many electronic deep neural networks do[47]. $l_{OP}$ is usually composed of an $l_1$- or $l_2$- norm between the middle results and the ideal output; $\alpha, \beta, \gamma$ are constant coefficients that balance these losses.

We use the optical fields in the fibers to characterize the high-dimensional latent values in the optical networks. The all-optical PED encodes the information in both the amplitude and phase domain. During training, the phase and amplitude are calculated based on the optical field in front of the coupling lens array with corresponding complex coupling coefficients. For experimental convenience, we also establish the model of optoelectronic PED. It allows the PED to achieve comparable performance with only one phase modulator. The output of each diffractive layer is captured with a sensor and the measured intensity is fed back to reuse the diffractive system. In this way, the optoelectronic PED achieves nonlinearity from the sensor but loses phase information to encode. Fig. S5 presents the comparison of the performance between the all-optical and optoelectronic PED on both general and data-specific modes. The all-optical and optoelectronic PED achieves tantamount results in the reconstruction fidelity and noise resistance. Additionally, the latency of this reusing architecture in optoelectronic PED has been proved ultra-low[30]. Therefore, the optoelectronic PED provides a convenient way to validate the all-optical PED. More details for the experimental implementation in Fig. 3-4 are included in the *Experimental set-up*.

During training, the proposed OLS complies with the assumption of a biased Gaussian distribution because the intensity of the optical signals is nonnegative. The deviation is set as a constant instead of learning during training to mimic the real fiber communication situations. Adjusting the deviation to different values helps the PED handle different noise levels.

The phase masks for the general and data-specific modes are shown in Fig. S6. The ablation analysis of diffractive distances and the layer number of the decoder for all-optical PED are included in Fig. S7. The growth of the performance slows when the layer number of the decoder exceeds two. While both the general and data-specific mode shows the same best diffractive distance as 15cm between the layers.

**Experimental set-up**
The system design and experimental set-up are shown in Fig S2. We use a single-mode 532 nm laser (Changchun New Industries Optoelectronics Tech. Co., Ltd., MGL-III-532-200mW) to generate the collimating incident light. A cascaded beam expander system is employed to expand the diameter of the beam to ~25 mm. We use an amplitude-modulation spatial light modulator (SLM) (HOLOEYE Photonics AG, HES6001) to generate the grayscale input and a phase-only SLM (Meadowlark Optics, Inc. P1920-400-800-PCIE) as the diffractive layer in the optical neural network. It has a frame rate of up to 714 Hz. A 4f-system is placed between the two SLMs to conjugate the input image onto the phase-modulation plane. The diffractive distance between the second SLM and the sensor is 300 mm. We use a scientific complementary metal-oxide semiconductor (Tucsen Photonics Co., Ltd., Dhyana 400BSI) to measure the intensity output from the optical neural network. In the optoelectronic PED, the multilayer network is realized by iteration with this system. The coupling of the fiber is measured in sequence by one fiber to mimic the effect of the array (Fig. S2c). The long-distance transmission noise is simulated according to the SNR as modeled in the Supplementary Information. We use the single-mode fiber (Daheng Optics, DH-FSM450-FC) at 532 nm.

**Adaptive training of the PED**
We employ adaptive training to correct some of the errors in the experimental system. By re-training the last few diffractive layers with experimental middle outputs over a small scale of the training dataset, adaptive training achieves distinct improvement on the test dataset. Because of the leverage of reconfigurable equipment in the set-up, we can easily change the weight in the PED after adaptive training.

*General mode:* Because the training and testing datasets are the same and the scale is small, we go through the whole dataset (512 instances). The noise is randomly added in each training epoch. Adaptive training fine-tunes the decoder masks for several epochs over data with random noise.

*Data-specific mode:* We experimentally test the output of the encoder on a small part of the training dataset (~3%) and fine-tune the decoder with this small amount of training data.

**Modeling of the coupling in the PED**
The corresponding lenses in the micro lens array divide the light field output from the encoder into $m$ areas, where $m$ is the dimensionality of the OLS. Each lens Fourier transfers the corresponding area to its frequency domain on its Fourier plane. Given single-mode fibers with a proper numerical aperture (NA), only the low frequencies are coupled into the fibers. The light in the fiber is a combination of light with different incident angles with corresponding complex coupling coefficients. The coupling efficiency is based on the data because the nonplanar wavefront affects the coupling. Figure S8a, b gives an example of the modulated light field on the front focal plane and the results after coupling by the lens array. Light with different incident angles has different complex coupling coefficients.

In this work, we simulate the complex coupling coefficients at different incident angles with the finite-difference time-domain (FDTD) method (Fig. S8c). In this way, any complicated incident wavefront can be decomposed into components from different angles to add up.

We take the vertical incidence as an example to verify the FDTD simulation with theoretical predictions. We use the Gaussian approximation of the zeroth-order Bessel function. The electric field of the fundamental mode on the edge is

$$E_{ff} = \frac{2}{\sqrt{\pi}\omega_0} \exp[-(\frac{r}{\omega_0})^2],[48]$$

where $r$ is the radial coordinate and $\omega_0$ is the radius of the single-mode field. The coupling efficiency is usually defined as

$$\eta = \frac{|\iint E_{if}^* E_{ff} ds|^2}{\sqrt{\iint |E_{if}|^2 ds \cdot \iint |E_{ff}|^2 ds}},[48]$$

where $E_{if}$ is the electric field of the incident light on the back focal plane and $E_{ff}$ is the electric field on the fiber edge. The FDTD simulation corresponds well with the theoretical result in this example.

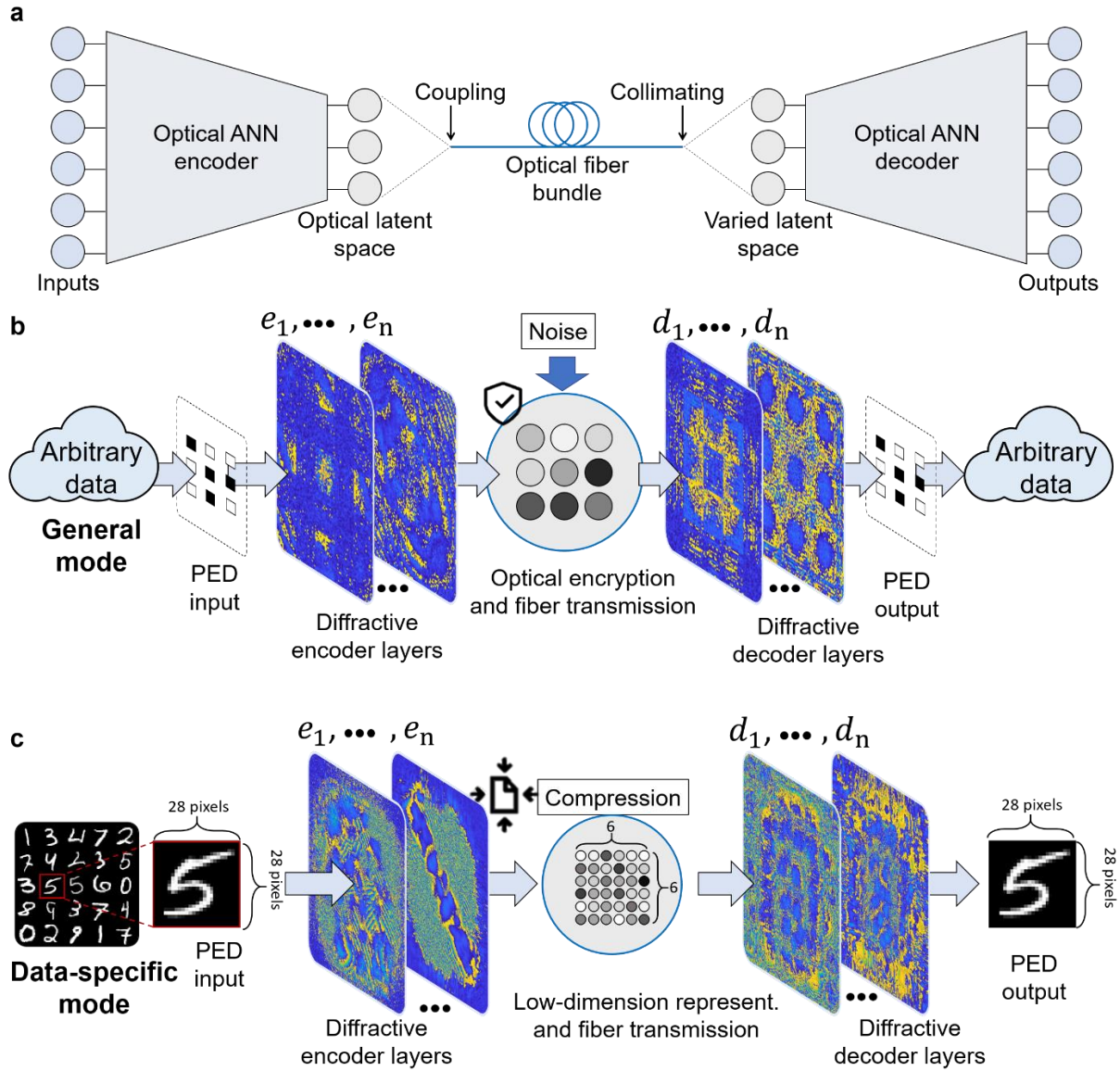

**Fig. 1. OFC using the PED neural network architecture. a**. The PED, implemented with photonic VAE, is used for end-to-end designing of an OFC system, which comprises the optical neural network of an encoder and a decoder and transmits the optical latent space with an optical fiber bundle. **b**. The general mode: the PED allows optically encrypting the digital inputs and reducing the transmission error with a non-data-specific pre-trained coding architecture. **c**. The data-specific mode: the PED compresses the grayscale inputs into low-dimensional representations with the encoder and reconstructs with its decoder for high throughput OFC.

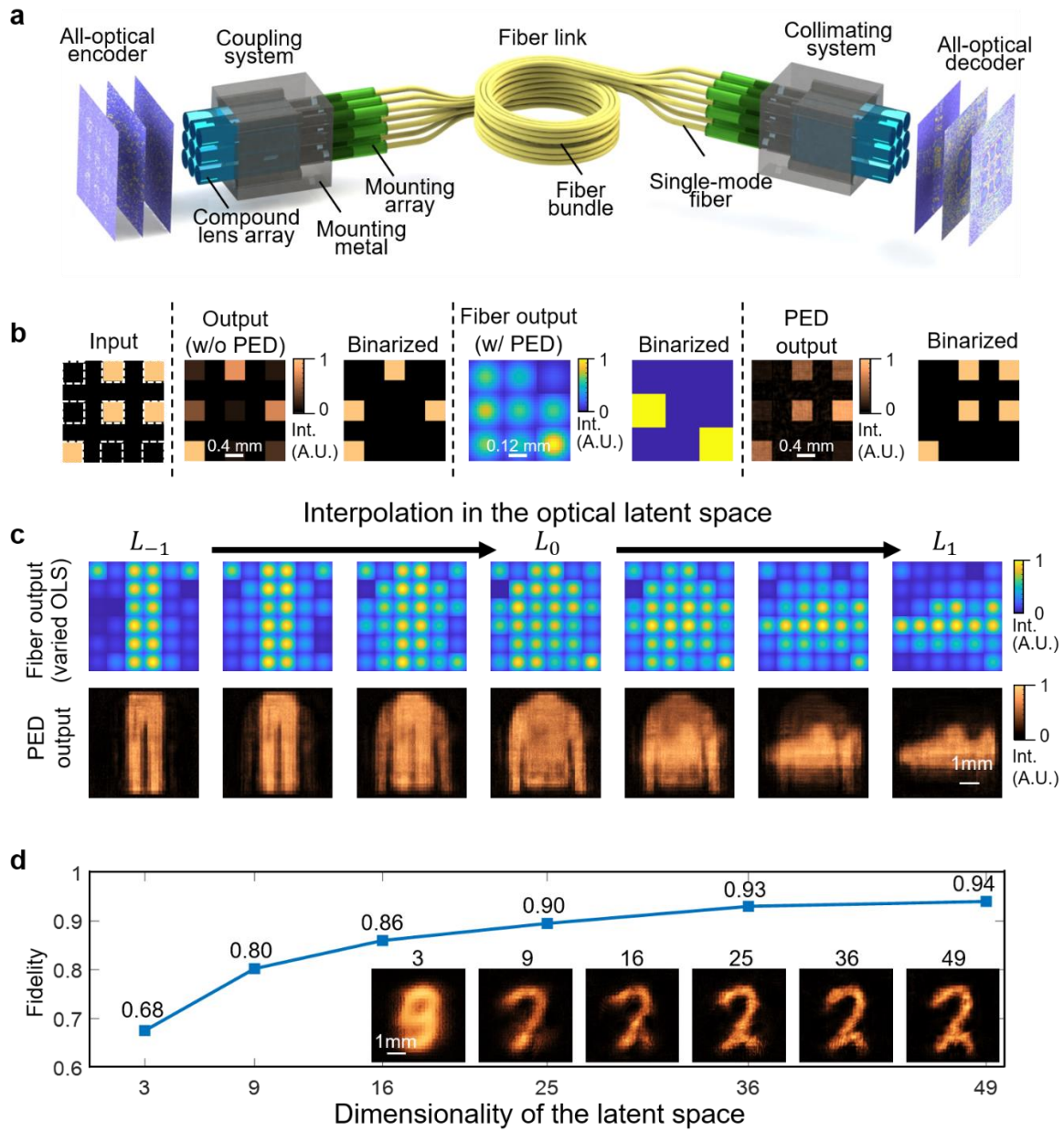

**Fig. 2. Modeling of the PED. a**. The system design of the PED. It couples the light-field output by the encoder into fiber bundles and decodes the information by the decoder all-optically. **b**. the OLS in the general mode: the encoder encrypts the inputs into their OLS representations, where the decoder can correct the bit error induced by the transmission noise and reduce the bit error ratio. **c**. the OLS in the data-specific mode: the OLS representations of inputs (the fiber outputs) are imposed by a continuous Gaussian distribution function, where the interpolation is uniform, while the reconstructions assemble around the original items. **d**. The reconstruction quality (fidelity) improves with the increasing of the OLS dimensionality rapidly when the compression ratio is large and turns gradually when the OLS dimensionality is adequate. Examples of the reconstructed digit '2' at different OLS dimensionality is displayed.

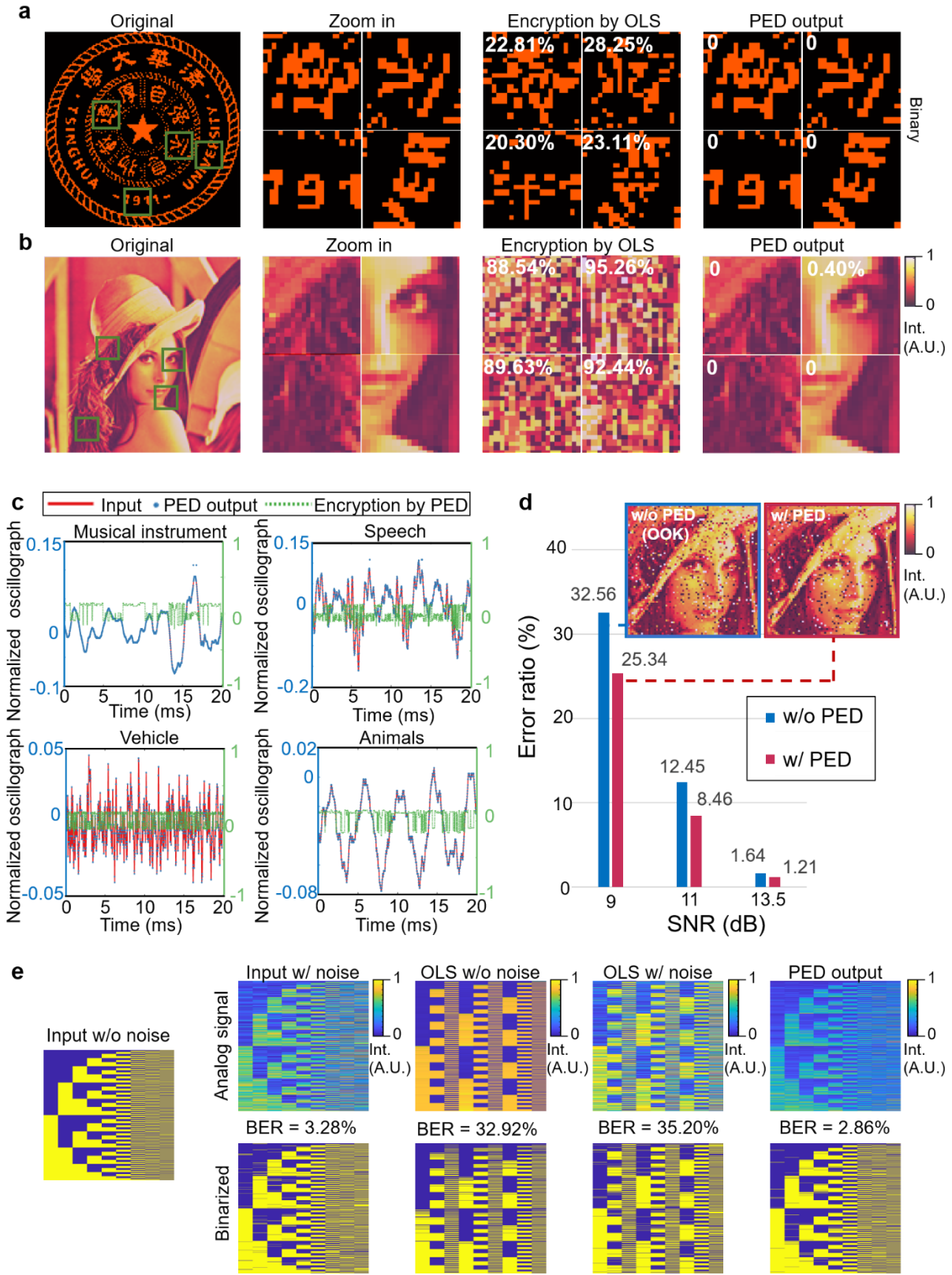

**Fig. 3. Experimental results of PED for the general mode. a-c.** Encrypted transmission of binary images, grayscale images and audios. All successfully distort the information in OLS and reconstruct after decoders. **d**. The error ratio of the PED under different transmission noise levels. We encode the grayscale image of 'Lena' into binary bits and transmit it with simulated noise. PED reduces the ER by up to 32.05% (relatively) compared with OOK (one of the existing methods that are most robust to transmission noise). **e**. Denoising results of the general mode on all 512 instances. We transmit all coding instances (512 images) to calibrate and test the global encryption and denoising capability of PED at a noise level of 9 dB. The same level of noise is added in both the OOK and the OLS. PED encrypts the information with the bit error ratio (BER) of 35.20% and reconstructs it to the BER of 2.86%. After coding, it reduces the BER by 12.8% (relatively) than OOK.

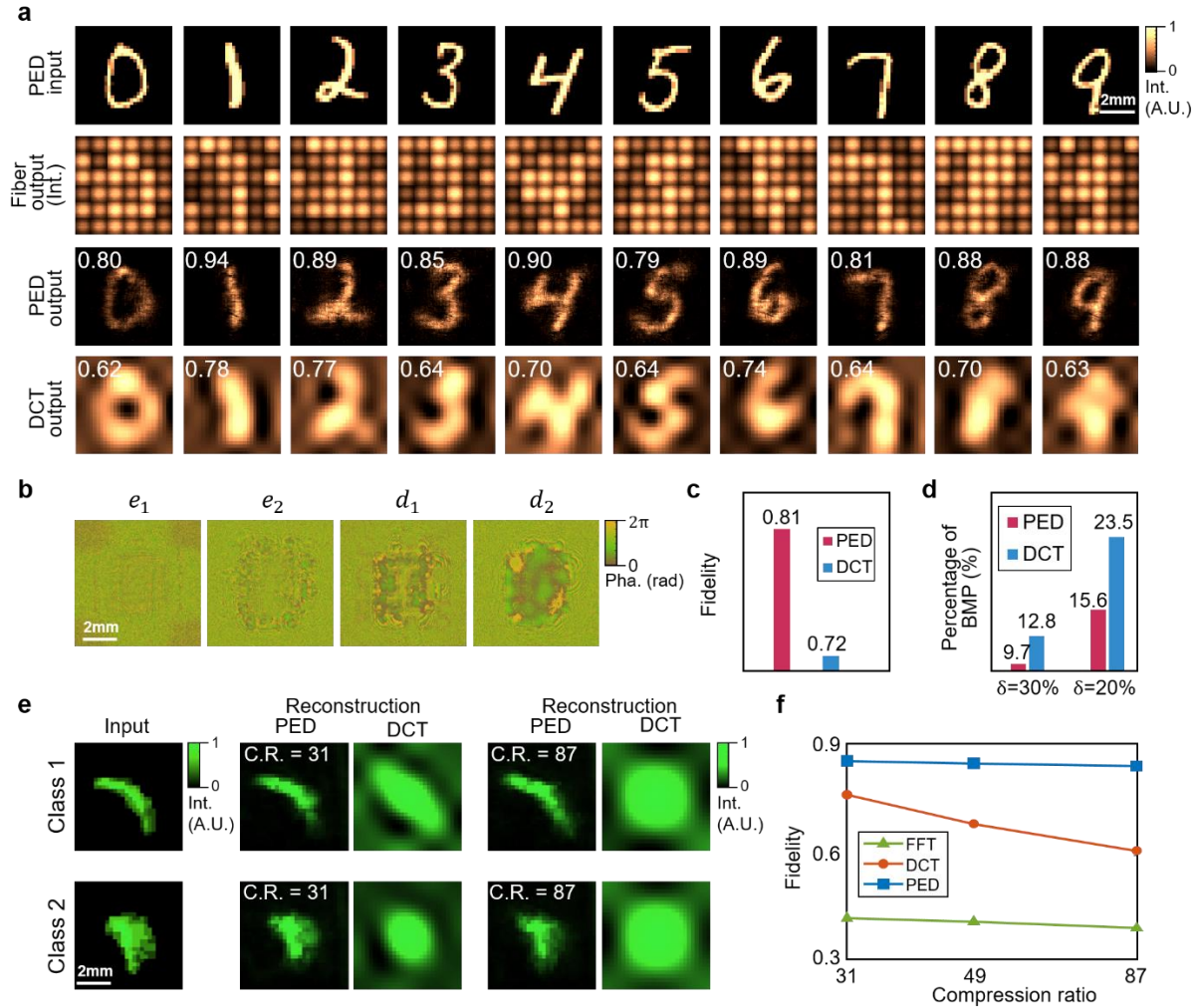

**Fig. 4. Experimental results of PED for the data-specific mode. a.** The inputs, i.e., handwritten digits from 0 to 9, 28× 28 grayscale pixels each (the first row), are optically transformed by the encoder and transmitted using their OLS representations. The OLS with a dimensionality of 6×6 is obtained by fiber coupling the optical field of different subregions on the encoder output plane, where the transmission noise is included as the variation of the OLS representations (the second row). Both amplitude noise (24 dB) and phase noise in transmission are simulated in the fiber link. The optical fields of fiber outputs are optically decoded by the decoder to reconstruct the input handwritten digits (the third row), demonstrating superior performance than DCT compression under the same compression ratio (the fourth row). Their reconstruction fidelities are labeled on each corner. **b.** The phase masks of the diffractive layers of the PED in this task. $e_n$ $(n = 1,2)$ is the $n$th layer of the encoder and $d_n$ $(n = 1,2)$ is the $n$th layer of the decoder. **c-d.** The reconstruction fidelity and percentage of bad matching pixels (BMP) of the PED and DCT averaged over MNIST test dataset (10k images). **e.** The reconstruction of the PED compared with DCT at different compression ratios over the projected adrenal CT images from MedMNIST[42]. C.R.: compression ratios. **f.** The reconstruction fidelity of the PED, DCT and FFT at different compression ratios. The DCT surpasses the other two mainstream methods significantly over all demonstrated compression ratios.